\documentclass[conference]{IEEEtran}
\usepackage{amssymb}
\usepackage{graphicx}

\ifCLASSINFOpdf
\else
\fi
\begin{document}
%
\title{A Discrete Time Markov Chain Model for High Throughput Bidirectional Fano Decoders}



%
\author{\IEEEauthorblockN{Ran Xu\IEEEauthorrefmark{1},
Graeme Woodward\IEEEauthorrefmark{2},
Kevin Morris\IEEEauthorrefmark{1} and
Taskin Kocak\IEEEauthorrefmark{1}}\\
\IEEEauthorblockA{\IEEEauthorrefmark{1}Centre for Communications Research, Department of Electrical and Electronic Engineering\\ University of Bristol, Bristol, UK}

\IEEEauthorblockA{\IEEEauthorrefmark{2}Telecommunications
Research Laboratory (TRL), Toshiba Research Europe Limited, 32 Queen Square, Bristol, UK}}



\maketitle

\begin{abstract}
The bidirectional Fano algorithm (BFA) can achieve at least two times decoding throughput compared to the conventional unidirectional Fano algorithm (UFA). In this paper, bidirectional Fano decoding is examined from the queuing theory perspective. A Discrete Time Markov Chain (DTMC) is employed to model the BFA decoder with a finite input buffer. The relationship between the input data rate, the input buffer size and the clock speed of the BFA decoder is established. The DTMC based modelling can be used in designing a high throughput parallel BFA decoding system. It is shown that there is a trade-off between the number of BFA decoders and the input buffer size, and an optimal input buffer size can be chosen to minimize the hardware complexity for a target decoding throughput in designing a high throughput parallel BFA decoding system.\\
\end{abstract}




%
\IEEEpeerreviewmaketitle

\section{Introduction}
Sequential decoding is one method for decoding convolutional codes [1]. Compared to the well-known Viterbi algorithm, the computational effort of sequential decoding is adaptive to the signal-to-noise-ratio (SNR). When the SNR is relatively high, the computational complexity of sequential decoding is much lower than that of Viterbi decoding. Additionally, sequential decoding can decode very long constraint length convolutional codes since its computational effort is independent of the constraint length. Thus, a long constraint length convolutional code can be used to achieve a better error rate performance.
There are mainly two types of sequential decoding algorithms which are known as the Stack algorithm [2] and the Fano algorithm [3]. The Fano algorithm is more suitable for hardware implementations since it does not require extensive sorting operations or large memory as the Stack algorithm [4][5].

High throughput decoding is of research interest due to the increasing data rate requirement. The baseband signal processing is becoming more and more power and area hungry. For example, to achieve the required high throughput, the WirelessHD specification proposes simultaneous transmission of eight interleaved codewords, each encoded by a convolutional code [6]. It is straightforward to use eight parallel Viterbi decoders to achieve multi-Gbps decoding throughput. Since sequential decoding has the advantage of lower hardware complexity and lower power consumption compared to Viterbi decoding [4][5], we are motivated to consider the usage of sequential decoding in high throughput applications when the SNR is relatively high. In a practical implementation of a sequential decoder, an input buffer is required due to the variable computational effort of each codeword. The contribution of this work is that the bidirectional Fano decoder with an input buffer was modelled by a Discrete Time Markov Chain (DTMC) and the relationship between the input data rate, the input buffer size and the clock speed of the BFA decoder was established. The trade-off between the number of BFA decoders and the input buffer size in designing a high throughput parallel BFA decoding system was also presented.

The rest of the paper is organized as follows. In Section II, the bidirectional Fano algorithm is reviewed and the system model is given. The BFA decoder with an input buffer is analyzed by queuing theory in Section III, and the simulation results are presented in Section IV. Section V is about choosing the optimal input buffer size in designing a parallel BFA decoding system, and the conclusions are drawn in Section VI.

\section{System Model for BFA Decoder}
\subsection{Bidirectional Fano Algorithm}
In the conventional unidirectional Fano algorithm (UFA), the decoder starts decoding from state zero. During each iteration of the algorithm, the current state may move forward, move backward, or stay at the current state. The decision is made based on the comparison between the threshold value and the path metric. If a forward movement is made, the threshold value needs to be tightened. If the current state cannot move forward or backward, the threshold value needs to be loosened. A detailed flowchart of the Fano algorithm can be found in [1]. In [7], a bidirectional Fano algorithm (BFA) was proposed, in which there is a forward decoder (FD) and a backward decoder (BD) working in parallel. Both the FD and the BD decode the same codeword from the start state and the end state in the opposite direction simultaneously. The decoding will terminate if the FD and the BD merge with each other or reach the other end of the code tree. Compared to the conventional UFA, the BFA can achieve a much higher decoding throughput due to the reduction in computational effort and the parallel processing of the two decoders. A detailed discussion on the BFA can be found in [7].

\subsection{System Model}

Since the computational effort of sequential decoding is variable, an input buffer is used to accommodate the codewords to be decoded. The system model for a BFA decoder with an input buffer is shown in Fig.\,1. It is assumed that there is continuous data stream input to the buffer whose raw data rate is $R_d$ bps. The length of the input buffer is $B$, which means that it can accommodate up to $B$ codewords, in addition to the one the decoder works on. The clock frequency of the BFA decoder is $f_{\it{clk}}$ Hz and it is assumed that the BFA decoder can execute one iteration per clock cycle. In the BFA decoding, the number of clock cycles to decode one codeword follows the Pareto distribution, and the Pareto exponent is a function of the SNR and the code rate. A higher SNR or a lower code rate results in a higher Pareto exponent [7]. As shown in Fig.\,1, there is an overflow notification from the input buffer to the BFA decoder. The occupancy of the input buffer is observed and the currently decoded codeword will be erased if the input buffer gets full. As a result, the total number of codewords consists of the following:
\begin{equation}
N_{\it{total}}=N_{\it{decoded}}+N_{\it{erased}}.
\end{equation}
In order to evaluate the performance of a BFA decoder affected by the introduced parameters such as $R_d$, $f_{\it{clk}}$ and $B$, a metric called failure probability ($P_f$) is defined as follows:
\begin{equation}
P_f=\frac{N_{\it{erased}}}{N_{\it{total}}}=\frac{N_{\it{erased}}}{N_{\it{decoded}}+N_{\it{erased}}},
\end{equation}
where $P_f$ is similar to the frame error rate ($P_F$) which is caused by the decoding errors. The total frame error rate is:
\begin{equation}
P_t=P_f+P_F.
\end{equation}
In designing the system, $R_d$, $f_{\it{clk}}$ and $B$ need to be chosen properly to ensure that:
\begin{equation}
P_t \approx P_F.
\end{equation}
In this paper, $P_f=0.01 \times P_F$ is adopted as the target failure probability ($P_{\mathit{target}}$). How to choose $R_d$, $f_{\it{clk}}$ and $B$ to make a BFA decoder achieve $P_{\mathit{target}}$ will be discussed next.

\begin{figure}[tb]
\centering
\includegraphics[scale=0.3]{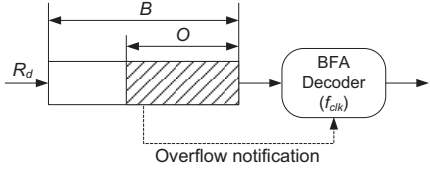}
\caption{System model for BFA decoder with overflow notification from the input buffer}
\label{fig1:}
\end{figure}


\section{DTMC based Modelling on BFA Decoder}
The effect of the input buffer has been investigated for iterative decoders such as Turbo decoder [8] and LDPC decoder [9]-[11]. The non-deterministic decoding time nature of the BFA is similar to that of Turbo decoding and LDPC decoding. A modelling strategy similar to that introduced in [11] is used to analyze the BFA decoder with input buffer.

The relationship between the input data rate ($R_d$), the input buffer size ($B$) and the clock speed of the decoder ($f_{\mathit{clk}}$) can be found via simulation. Another way to analyze the system is to model it based on queuing theory. The BFA decoder with an input buffer can be treated as a \textbf{D/G/1/B} queue, in which \textbf{D} means that the input data rate is deterministic, \textbf{G} means that the decoding time is generic, \textbf{1} means that there is one decoder and \textbf{B} is the number of codewords the input buffer can hold. The state of the BFA decoder is represented by the input buffer occupancy ($O$) when a codeword is decoded, which is measured in terms of branches stored in the buffer. $O(n)$ and $O(n+1)$ have the following relationship:
\begin{equation}
O(n+1)=O(n)+[T_s(n) \cdot R_{d}-L_{f}],
\end{equation}
where $O(n+1)$ is the input buffer occupancy when the $n^{th}$ codeword is decoded, $T_s(n)$ is the decoding time of the $n^{th}$ codeword by the BFA decoder and $L_f$ is the length of a codeword in terms of branches. $[x]$ denotes the operation to get the nearest integer of $x$. The speed factor of the BFA decoder is defined as the ratio between $f_{\it{clk}}$ and $R_d$ [1]:
\begin{equation}
\mu=\frac{f_{\it{clk}}}{R_d}.
\end{equation} 
If $f_{\mathit{clk}}$ is normalized to 1, Eq.\,(5) can be changed to:
\begin{equation}
O(n+1)=O(n)+[\frac{T_s(n)}{\mu}-L_{f}].
\end{equation}
The state of the input buffer at time $n+1$ is only decided by the state at time $n$ and the decoding time $T_s(n)$. At the same time, $T_s(n)$ and $T_s(n+1)$ are \emph{i.i.d.}. As a result, the state of the input buffer is a Discrete Time Markov Chain (DTMC). $T_s(n)$ follows the Pareto distribution for the BFA decoding and is in the unit of \emph{clock cycle/codeword}. The following equation can be used to describe the Pareto distribution:
\begin{equation}
\mathit{Prob}(T_s>T) \approx A \cdot (\frac{T}{T_{\mathit{min}}})^{-\beta},
\end{equation}
where $T_{\mathit{min}}$ is the minimum decoding time which is $L_f$ clock cycles in the considered model. The Pareto exponent $\beta$ is a function of the SNR and the code rate. Fig.\,2 shows the simulated and approximated (based on Eq.\,(8)) Pareto distributions for both the UFA and the BFA at $E_b/N_0$=4dB and 5dB. It can be seen that as the SNR increases, the Pareto exponent increases, and for the same SNR the BFA has a higher Pareto exponent compared to the UFA. The simulated Pareto distribution of $T_s$, which is more accurate compared to the approximated distribution based on Eq.\,(8), will be used in the following analysis. The difference between $O(n+1)$ and $O(n)$ is defined as:
\begin{equation}
\Delta(n)=O(n+1)-O(n)=[\frac{T_s(n)}{\mu}-L_{f}].
\end{equation}
Fig.\,3 shows that the total number of states of the input buffer with size $B$ is:
\begin{equation}
\Omega=B \cdot L_f.
\end{equation}
The state transition diagram is shown in Fig.\,4. As a result, the state transition probability matrix of the input buffer is:
\begin{equation}
\matrix{P_T}=\left( \begin{array}{cccc}
P_{11} & P_{12} & \cdots & P_{1\Omega}\\
P_{21} & P_{22} & \cdots & P_{2\Omega}\\
\vdots & \vdots & \ddots & \vdots\\
P_{\Omega 1} & P_{\Omega 2} & \cdots & P_{\Omega\Omega}\\
\end{array}
\right),
\end{equation} 
where $P_{ij}$ is the state transition probability from $S_i$ to $S_j$, which can be calculated as follows:
\begin{equation}
P_{ij}=\left\{ \begin{array}{rc}
\sum_{k=\Delta_{min}}^{-(i-1)} p_{\Delta_{+k}}, & j=1\\\\
p_{\Delta_{+(j-i)}}, & 1<j<\Omega\\\\
1-\sum_{k=1}^{\Omega-1} P_{ik}, & j=\Omega
\end{array}
\right.,
\end{equation}
where $p_{\Delta_{+w}}=\mathit{Prob}(\Delta=w)$ and $\Delta_{min}=[\frac{min(T_s)}{\mu}-L_f]$. The value of $p_{\Delta_{+w}}$ can be estimated from the simulated distribution of $T_s$ as shown in Fig.\,2. The initial state probability ($n$=0) of the input buffer is:
\begin{equation}
\pi(0)=(\pi_1(0),\pi_2(0),\ldots,\pi_\Omega(0))=(1,0,\ldots,0).
\end{equation}
The steady state probability of the input buffer is then:
\begin{equation}
\Pi=\lim_{n \rightarrow \infty} \pi(n)=\lim_{n \rightarrow \infty} \pi(0) \cdot P_{T}^n.
\end{equation} 
The failure probability of the decoder can be calculated by:
\begin{equation}
P_f=\sum_{i=1}^{\Omega} \Pi(i) \cdot p_{\Delta_{\Omega-i}}^{+},
\end{equation} 
where $p_{\Delta_{\Omega-i}}^{+}=\mathit{Prob}(\Delta>\Omega-i)$. The mean buffer occupancy can be calculated by:
\begin{equation}
O_{\mathit{mean}}=\frac{\sum_{i=1}^{B} i\cdot \sum_{j=1}^{L_f} \Pi((i-1) \cdot L_f+j)}{B} \times 100\%.
\end{equation}

\begin{figure}[t]
\centering
\includegraphics[height=60mm,width=85mm]{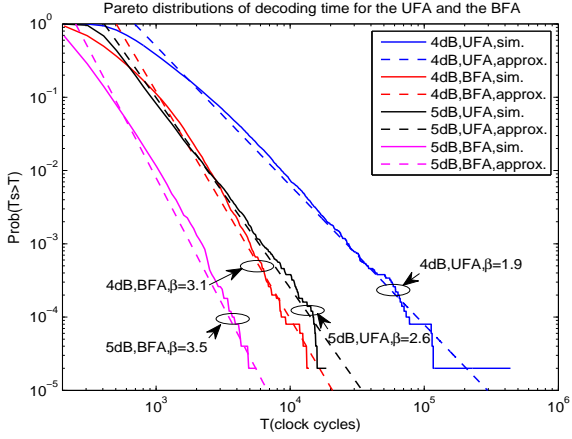}
\caption{Simulated and approximated Pareto distributions for the UFA and the BFA at $E_b/N_0$=4dB and 5dB. The code rate is $R$=1/3.}
\label{fig2:}
\end{figure}

\begin{figure}[tb]
\centering
\includegraphics[scale=0.4]{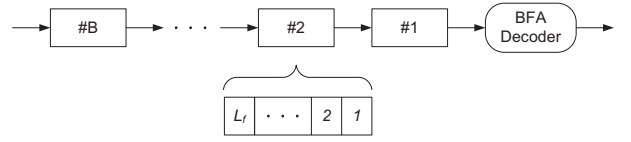}
\caption{BFA decoder with finite input buffer}
\label{fig3:}
\end{figure}

\begin{figure}[tb]
\centering
\includegraphics[scale=0.2]{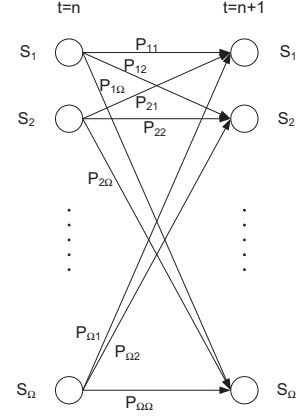}
\caption{Illustration of state transition}
\label{fig4:}
\end{figure}

\section{Simulation Results}
Firstly, the semi-analytical results calculated by Eq.\,(15) are compared with the simulation results to validate the DTMC based modelling. The simulation setup is shown in Table 1. $E_b/N_0$=4dB was used as an example, at which $P_{\mathit{target}} \approx 10^{-3}$. The convolutional code in the simulation was the one used in the WirelessHD specification [6]. The input buffer size $B$ in the simulation takes the buffer within the BFA decoder into account. It can be seen from Fig.\,5 that the semi-analytical results are quite close to the simulation results for both the UFA decoder and the BFA decoder, which means that the DTMC based modelling is accurate. For the input buffer size of $B$=10, the working speed factors of the UFA decoder and the BFA decoder are about $\mu$=14 and $\mu$=3.6, respectively. There is about 290\% decoding throughput improvement by using the BFA decoder compared to the UFA decoder. If the input buffer size increases to $B$=25, the working speed factors will become about $\mu$=8.7 and $\mu$=2.9, respectively, resulting in about 200\% decoding throughput improvement. As long as the distribution of $T_s$ is known, $P_f$ can be easily obtained for different values of speed factor and input buffer size. Simulation time can be greatly saved if the target $P_f$ is very low (at high SNR) by using the DTMC based modelling. How to use the DTMC based modelling in designing a high throughput parallel BFA decoding system will be shown in the next section.

\begin{table}[tb]
\centering
\caption{Simulation setup}
\label{tab1:}
\begin{tabular}{|c||c|}
\hline
Code rate ($R$) & $1/3$\\
\hline
Generator polynomials & $g_0=133_8$, $g_1=171_8$, $g_2=165_8$\\
\hline
Constraint length ($K$) & 7\\
\hline
Branch metric calculation & 1-bit hard decision with Fano metric\\
\hline
Threshold adjustment value ($\delta$) & 2\\
\hline
Modulation & BPSK\\
\hline
Channel & AWGN\\
\hline
Information length ($L$) & 200 bits\\
\hline
Codeword length ($L_f$) & $L+K-1=206$ branches\\
\hline
\end{tabular}
\end{table}

The input buffer occupancy distribution for the BFA decoder with $B$=10 at different speed factors is shown in Fig.\,6, which was obtained from Eq.\,(14). The mean buffer occupancy in percentage calculated by Eq.\,(16) is shown in Fig.\,7. For both $B$=10 and $B$=25 whose working speed factors are about 3.6 and 2.9, the mean buffer occupancies are about 17\% and 25\%, respectively. The decoding delay for $B$=25 is slightly higher than that for $B$=10, while the decoding throughput for $B$=25 is higher than that for $B$=10 as shown in Fig.\,5.

\begin{figure}[tb]
\centering
\includegraphics[height=60mm,width=85mm]{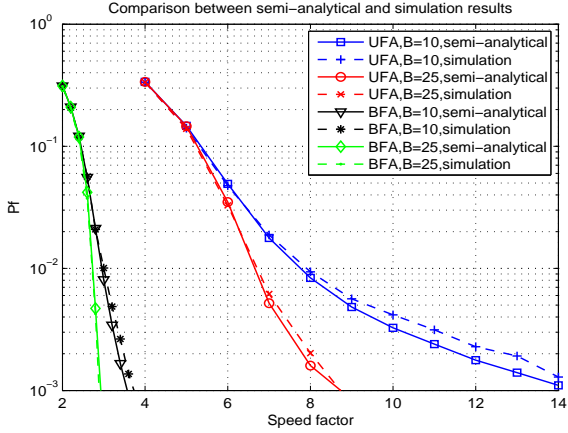}
\caption{Comparison between semi-analytical and simulation results ($P_f$ vs $\mu$) for UFA and BFA at $E_b/N_0$=4dB}
\label{fig5:}
\end{figure}

\begin{figure}[t]
\centering
\includegraphics[height=60mm,width=85mm]{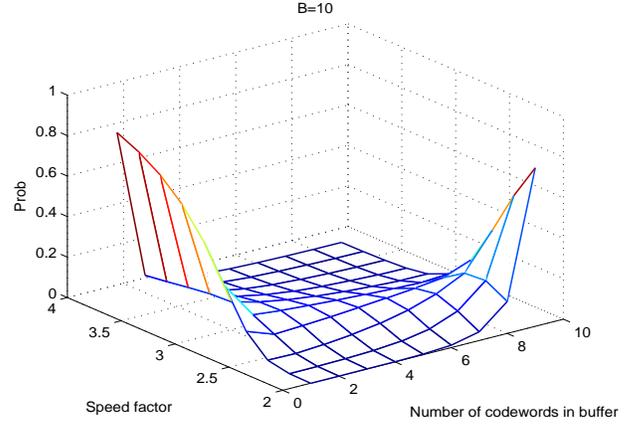}
\caption{Buffer occupancy distribution for BFA decoder at $E_b/N_0$=4dB when $B$=10}
\label{fig6:}
\end{figure}

\begin{figure}[tb]
\centering
\includegraphics[height=60mm,width=85mm]{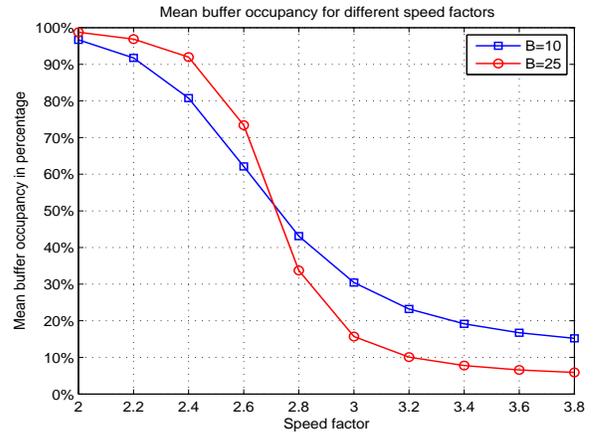}
\caption{Mean buffer occupancy for BFA decoder at $E_b/N_0$=4dB when $B$=10 and $B$=25}
\label{fig7:}
\end{figure}

\section{Input Buffer Size in Parallel BFA Decoding}
Unlike the Viterbi decoder, it is difficult to use pipelining in designing a high throughput BFA decoder due to the irregular decoding operations and the variable computational effort. Parallel processing is a promising strategy to achieve high throughput BFA decoding at multi-Gbps level. In order to achieve a specific decoding throughput, a number of BFA decoders ($N_{\mathit{decoder}}$) may need to be paralleled (as shown in Fig.\,8) if a single BFA decoder cannot achieve the target average decoding throughput:
\begin{equation}
T_{\mathit{target}}=N_{decoder} \cdot R_d(B),
\end{equation}
where $R_d$ is a function of the input buffer size $B$. The total area of the parallel BFA decoders is:
\begin{eqnarray}
\mathcal{A}_{\mathit{total}} & = & \mathcal{A}_{\mathit{decoder}}+\mathcal{A}_{\mathit{buffer}} \nonumber \\
          & = & N_{\mathit{decoder}} \cdot \mathcal{A}_{\mathit{BFA}}+N_{\mathit{decoder}} \cdot B \cdot \mathcal{A}_{B}.
\end{eqnarray}
If the area ratio between a BFA decoder ($\mathcal{A}_{\mathit{BFA}}$) and an input buffer which can hold one codeword ($\mathcal{A}_{B}$) is $\alpha=\mathcal{A}_{\mathit{BFA}} / \mathcal{A}_{B}$, Eq.\,(17) will become:
\begin{equation}
T_{\mathit{target}}=\frac{\mathcal{A}_{\mathit{total}}}{\mathcal{A}_{\mathit{BFA}}+B \cdot \mathcal{A}_{B}} \cdot R_d(B)=\frac{\mathcal{A}_{\mathit{total}}}{\mathcal{A}_{B}} \cdot \frac{R_d(B)}{\alpha+B}.
\end{equation}
It can be seen from Eq.\,(19) that for a fixed $\mathcal{A}_{\mathit{total}}$ and $\mathcal{A}_B$, the decoding throughput of parallel BFA decoders changes with respect to the input buffer size $B$. The relationship between the input data rate $R_d$ and input buffer size $B$ is shown in Fig.\,9 which was obtained by the DTMC based modelling introduced in Section III. The clock speed of the BFA decoder is assumed to be $f_{\mathit{clk}}$=1GHz. The normalized throughput with respect to the maximum throughput for different $\alpha$ values is shown in Fig.\,10. The value of $\alpha$ depends on the technology used in hardware implementation. It can be seen from Fig.\,10 that there is an optimal choice of the input buffer size $B$ to maximize the decoding throughput for a fixed area constraint. For example if $\alpha$=16, the optimal choice of the input buffer size will be 10. Equivalently, in order to achieve a target decoding throughput, the optimal choice of the input buffer size can minimize the hardware area, which will be explained by the following example.\\\\ $\blacksquare$ \textit{Example} \\ If the target decoding throughput is $T_{\mathit{target}}$=1Gbps and two input buffer sizes $B_1$=5 and $B_2$=10 are used, according to Eq.\,(17) and Fig.\,9, the number of parallel BFA decoders required are:
\begin{equation}
N_1=6 \mbox{ and } N_2=4.
\end{equation}
When $B_1$=5 is used, the total area of the parallel BFA decoders will be:
\begin{equation}
\mathcal{A}_1=N_1 \cdot \mathcal{A}_{\mathit{BFA}}+N_1 \cdot B_1 \cdot \mathcal{A}_B.
\end{equation}
When $B_2$=10 is used, the total area of the parallel BFA decoders will be:
\begin{equation}
\mathcal{A}_2=N_2 \cdot \mathcal{A}_{\mathit{BFA}}+N_2 \cdot B_2 \cdot \mathcal{A}_B.
\end{equation}
If $\alpha$=16, the area reduction by using $B_2$=10 compared to $B_1$=5 will be:
\begin{eqnarray}
\eta & = & (\frac{\mathcal{A}_1}{\mathcal{A}_2}-1) \times 100 \% \nonumber \\ 
          & = & (\frac{N_1}{N_2} \cdot \frac{\alpha+B_1}{\alpha+B_2}-1) \times 100 \% \approx 20\%.
\end{eqnarray}

\begin{figure}[t]
\centering
\includegraphics[scale=0.42]{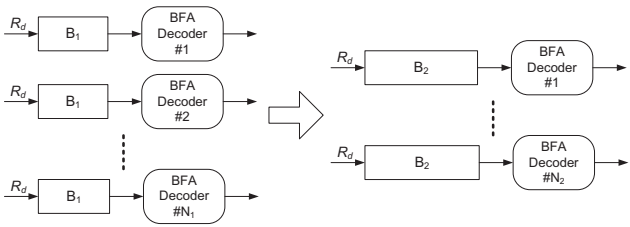}
\caption{Number of decoders vs input buffer size in parallel BFA decoding}
\label{fig8:}
\end{figure}

\begin{figure}[t]
\centering
\includegraphics[height=60mm,width=85mm]{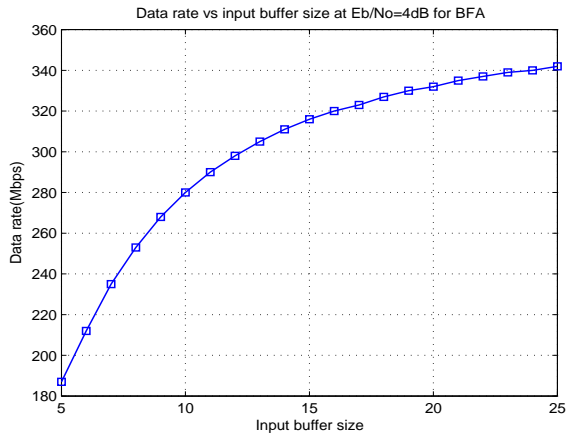}
\caption{Data rate vs input buffer size for BFA at $E_b/N_0$=4dB}
\label{fig9:}
\end{figure}

\begin{figure}[t]
\centering
\includegraphics[height=60mm,width=85mm]{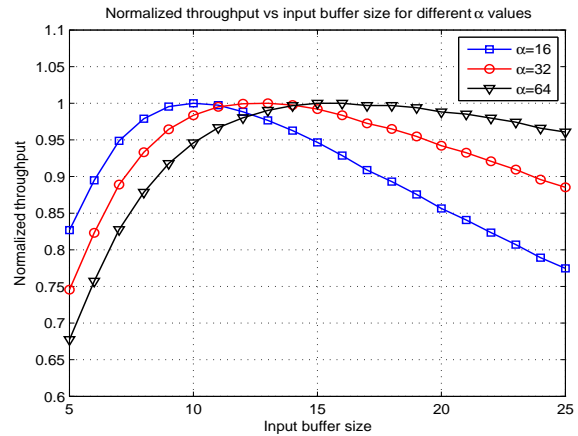}
\caption{Normalized throughput vs input buffer size for different $\alpha$ values at $E_b/N_0$=4dB}
\label{fig10:}
\end{figure}

\section{Conclusion}
In this paper, BFA decoder with input buffer was analyzed from the queuing theory perspective. The decoding system was modelled by a Discrete Time Markov Chain and the relationship between the input data rate, the input buffer size and the clock speed of the decoder was established. The working speed factor of the BFA decoder at each SNR can be easily found by the DTMC based modelling. The DTMC based modelling can be used in designing a high throughput parallel BFA decoding system. The trade-off between the number of BFA decoders and the input buffer size in designing a high throughput parallel BFA decoding system was discussed as well. It was shown that an optimal input buffer size can be found for a target decoding throughput under a fixed hardware area constraint.

\section*{Acknowledgment}
The authors would like to thank the Telecommunications
Research Laboratory (TRL) of Toshiba Research Europe Ltd.
and its directors for supporting this work.




\begin{thebibliography}{1}

\bibitem{key-1}
S. Lin and D. J. Costello, Jr., \textit{Error
Control Coding: Fundamentals and Applications}, 2nd ed. Upper Saddle River, NJ: Pearson Prentice-Hall, 2004.

\bibitem{key-2}
F. Jelinek, \textquotedblleft{}Fast sequential decoding using a stack,\textquotedblright{} \textit{IBM J. Res. Devel.}, vol. 13, pp. 675-685, Nov. 1969.

\bibitem{key-3}R. M. Fano, \textquotedblleft{}A heuristic discussion of probabilistic decoding,\textquotedblright{} \textit{IEEE Transactions on Information Theory}, vol. IT-9, no. 2, pp. 64-74, Apr. 1963.

\bibitem{key-4}R. O. Ozdag and P. A. Beerel, \textquotedblleft{}An
asynchronous low-power high-performance sequential decoder implemented
with QDI templates,\textquotedblright{} \textit{IEEE Transactions on Very Large Scale Integration (VLSI) Systems}, vol. 14, no. 9, pp. 975-985, Sep. 2006.

\bibitem{key-5}M. Benaissa and Y. Zhu, \textquotedblleft{}Reconfigurable hardware architectures for sequential and hybrid decoding,\textquotedblright{} \textit{IEEE Transactions on Circuits and Systems I: Regular Papers}, vol. 54, no. 3, pp. 555-565, Mar. 2007.

\bibitem{key-6}\textquotedblleft{}Wireless High-Definition (WirelessHD)\textquotedblright{}; http://www.wirelesshd.org/

\bibitem{key-7}R. Xu, T. Kocak, G. Woodward, K. Morris and C. Dolwin, \textquotedblleft{}Bidirectional Fano Algorithm for High Throughput Sequential Decoding,\textquotedblright{} \textit{IEEE Symp. on Personal, Indoor and Mobile Radio Communications (PIMRC)}, Tokyo, Japan, 2009.

\bibitem{key-8}A. Martinez and M. Rovini, \textquotedblleft{}Iterative decoders based on statistical multiplexing,\textquotedblright{} \textit{Proc. 3rd Int. Symp. on Turbo Codes and Related Topics}, pp. 423-426, Brest, France, 2003.

\bibitem{key-9}M. Rovini and A. Martinez, \textquotedblleft{}On the Addition of an Input Buffer to an Iterative Decoder for LDPC Codes,\textquotedblright{} \textit{Proc. IEEE 65th Vehicular Technology Conference, VTC2007-Spring}, pp. 1995-1999, Apr. 2007.

\bibitem{key-10}S. L. Sweatlock, S. Dolinar, and K. Andrews, \textquotedblleft{}Buffering Requirements for Variable Iterations LDPC Decoders,\textquotedblright{} \textit{Proc. Information Theory and Applications (ITA) Workshop}, pp. 523-530, 2008.

\bibitem{key-11}G. Bosco, G. Montorsi, and S. Benedetto, \textquotedblleft{}Decreasing the Complexity of LDPC Iterative Decoders,\textquotedblright{} \textit{IEEE Communications Letters}, vol. 9, no. 7, pp. 634-636, July 2005.

\end{thebibliography}
%

\end{document}